\documentclass{article}

\usepackage{arxiv}

\usepackage[utf8]{inputenc} 
\usepackage[T1]{fontenc}    
\usepackage{hyperref}       
\usepackage{url}            
\usepackage{booktabs}       
\usepackage{amsfonts}       
\usepackage{nicefrac}       
\usepackage{microtype}      
\usepackage{lipsum}
\usepackage{graphics} 
\usepackage{epsfig} 
\usepackage{mathptmx} 
\usepackage{amsmath} 
\usepackage{amssymb}  

\title{Design to Automate the Detection and Counting of Tuberculosis(TB) Bacilli}

\author{
  Dinesh Jackson Samuel \\
  Postdoctoral Researcher\\
  Faculty of Technology, Design and Environment\\
  Visual Artificial Intelligence Laboratory\\
  Oxford Brookes University \\
  \texttt{rsamuel@brookes.ac.uk, jacksoncse@gmail.com} \\
   \And
    Rajesh Kanna Baskaran \\
	Professor\\
	School of Computer Science and Engineering\\
	Vellore Institute of Technology\\
	Chennai, India \\
	\texttt{brajeshkanna@vit.ac.in} 
}

\begin{document}
\maketitle

\begin{abstract}
Tuberculosis is a contagious disease which is one of the leading causes of death, globally. The general diagnosis methods for tuberculosis include microscopic examination, tuberculin skin test, culture method, enzyme linked immunosorbent assay (ELISA) and electronic nose system. World Health Organization (WHO) recommends standard microscopic examination for early diagnosis of tuberculosis. In microscopy, the technician examines field of views (FOVs) in sputum smear for presence of any TB bacilli and counts the number of TB bacilli per FOV to report the level of severity. This process is time consuming with an increased concentration for an experienced staff to examine a single sputum smear. The examination demands for skilled technicians in high-prevalence countries which may lead to overload, fatigue and diminishes the quality of microscopy. Thus, a computer assisted system is proposed and designed for the detection of tuberculosis bacilli to assist pathologists with increased sensitivity and specificity. The manual efforts in detecting and counting the number of TB bacilli is greatly minimized. The system obtains Ziehl-Neelsen stained microscopic images from conventional microscope at $100 \times$ magnification and passes the data to the detection system. Initially the segmentation of TB bacilli was done using RGB thresholding and Sauvola’s adaptive thresholding algorithm. To eliminate the non-TB bacilli from coarse level segmentation, shape descriptors like area, perimeter, convex hull, major axis length and eccentricity are used to extract only the TB bacilli features. Finally, the TB bacilli are counted using the generated bounding boxes to report the level of severity. 
\end{abstract}


\section{Introduction}
Tuberculosis is a contagious disease that spreads through Mycobacterium Tuberculosis and accounts for top causes of death, ranking above HIV/AIDS. According to WHO report on tuberculosis for year 2020, 10 million incident cases occur globally and 1.3 million people have died of it. Whereas, 5.6 million cases are found among men, 3.2 million cases among women and 1.2 million cases among children \cite{who}. People affected with HIV have the higher incident rate compared to the healthy ones.  India, China, Indonesia, Pakistan, Nigeria and South Africa account for 60 percent of new cases globally. In 2020, Multi Drug Resistant Tuberculosis (MDR-TB) had a sharp rise, with an estimation of 4,80,000 incident cases.

The diagnosis test for TB bacilli involves Polymerase Chain Reaction (PCR) test, Gene Xpert, chest X-ray, culture test, tuberculin skin test, and sputum smear microscopy examination \cite{Radzi2011ReviewOM}. Since the incidence rate is high in tuberculosis, World Health Organization (WHO) recommends microscopy examination as the early and effective technique for identification of tuberculosis bacilli. There are two different microscopes, optical microscope and fluorescence microscope, used for the detection of TB bacilli. The tuberculosis bacilli can be stained using dyes and thus are easy to characterize during microscopic examination. These Acid Fast Bacilli (AFB) can be stained using Auramine and Ziehl-Neelsen stains. Ziehl-Neelsen is used to stain TB bacilli for optical microscopy examination and Auramine is used in fluorescence microscope \cite{article}. Comparatively, an optical microscopy examination is less expensive and is used for early detection in many countries. The sputum is collected from the patient and are stained using Ziehl-Neelsen for smear preparation. Once the smear is prepared, skilled laboratory technician examines the slide for detecting Acid Fast Bacilli (AFB). On viewing over 100 microscopic fields, the technician gives the report for further diagnosis as, if there are more than 10 AFB per field of view then it is 3+, 1-10 AFB per 100 FOVs is 2+, 10-99 AFB in 100 FOVs is 1+, 1-9 AFB in 100 FOVs is scanty as shown in Table \ref{tab1} \cite{tbinc}.
\begin{table}[h!]
\centering

\begin{tabular}{|l|l|}
\hline
\textbf{{Occurrence of AFB bacilli in FOVs}}               & \textbf{{ Severity report}} \\ \hline
{ >10 AFB/field after examination of 20 fields}  & \textit{Positive,   3+}                \\ \hline
{1-10 AFB/field after examination of 50 fields} & \textit{Positive,   2+}                \\ \hline
{10-99 AFB/100 fields}                          & \textit{Positive,   1+}                \\ \hline
{1-9 AFB/100 fields}                            & \textit{Scanty}                        \\ \hline
{ No AFB per 100 fields}                         & \textit{Negative}                      \\ \hline
\end{tabular} \\
\vspace{0.3cm}
\caption{TB severity grading based on number of bacilli present in a field of view}
\label{tab1}
\end{table}
However, the manual microscopic examination in high prevalence countries may lead to overload, time consumption and fatigues which will diminish the sensitivity and specificity of the microscopy. Rapid detection and early identification is the goal in tuberculosis identification \cite{6630069}. Hence, the image processing and computer intelligence comes into the picture for accurate TB bacilli identification \cite{5735390d}. Costa et al. proposed the first computer aided AFB identification for conventional microscope \cite{article6}. The segmentation is done using the adaptive color threshold in RGB image, ensured by morphological operations to remove artifacts. The sensitivity rate is 76.65\% and has a false positive of 12\%. Sadaphal et al. used the Bayesian approach for segmenting the bacilli from the image \cite{article7}. Then the morphology features of the bacilli are extracted using axis ratio, eccentricity and area. Sotaquira et al. uses the YCbCr, lab color spaces for segmenting the bacilli and estimates the severity of infection by counting the total number of bacilli in each Field of View in an image \cite{5190506}. The sensitivity and specificity obtained is 90.9\% and ~ 100\%. Osman et al. used the Hybrid Multi-Layer Perceptron (HMLP) network and attained the accuracy of 99.82\% \cite{5605524}. The segmentation of the bacilli is done by Multi-Layer Perceptron (HMLP) network and the features are then passed into Modified Recursive Prediction Error (MRPE) method for training the HMLP network. Osman et al. proposed the moment invariant features and neural networks for AFB identification . Extreme Learning Machine (ELM) is used to train the HMLP network for better performance. CY color space and K-Mean is used for segmentation to attain the accuracy of 77.25\% \cite{5759878s}. Recently, Dinesh et. al. automated the complete Tuberculosis diagnosis system with a programmable microscopic stage and a human intelligence simulated recognition system for identifying infected field of views. The programmable framework defines the scanning direction for the specimen and a firmware is designed and developed which is used to control the microscopic stage, to enable customization of the scanning pattern without any human intervention during the complete course of screening \cite{https://doi.org/10.1002/jemt.23184}. Once the data is acquired, the frames are passed to a recognition system which uses deep leaning techniques. Here the domain adaptation and transfer learning are applied to the pre trained VGG and Inception V3 deepnets for representational learning \cite{samuel2019tuberculosis}, \cite{rajesh2018cybernetic}. 

The proposed method focuses on automating the AFB identification from the ZN stained microscopic image. Microscopic images are digitized using the camera and processed for AFB detection. The coarse and fine level segmentation provides more accuracy in segmenting the TB bacilli. RGB color thresholding is used to extract bacilli pixels for coarse segmentation followed by Sauvola’s local thresholding algorithm for fine segmentation \cite{inproceedings}. After segmentation, the shape descriptors like area, perimeter, convex hull perimeter, roughness, circularity and major axis length are calculated to extract the AFB bacilli. On detection of TB bacilli from the image, TB bacilli are counted to ensure the severity of the infection.

\section{Methodology}
The sputum is collected from the patient and stained using the Zeihl-Neelsen (ZN) stain. These stained specimens were obtained from Pondicherry Institute of Medical Sciences (PIMS), Pondicherry. The ZN stained smear is then examined by the technician for TB bacilli using bright field microscope. The bright field microscope has ocular lens, objective lens, light source and a microscopic stage to hold the specimen. To examine a ZN stained smear, the objective has to be set as 100x magnification for AFB identification. The TB bacilli appear to be red in a bluish background. There is also some non TB bacilli which are present in the smear. The challenge lies in detecting and eliminating those non TB bacilli before reporting. 

\begin{figure}[h!]
\centering
\includegraphics[width=5cm, height=12cm]{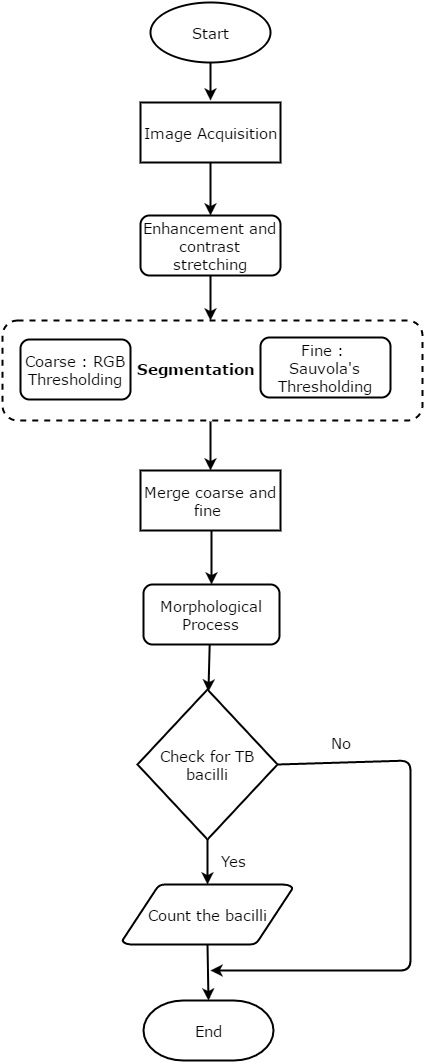}
\centering
\caption{Tuberculosis bacilli detection pipeline}
\label{fig1}
\end{figure}

Segmentation of TB bacilli is done based on color and shape descriptors. In Ziehl-Neelsen stain, bacilli which appear red are classified as TB bacilli using color descriptors. The shape descriptors include eccentricity, compactness, axis ratio, Hu moments and Zernike moments. The bacilli may appear in clusters or in overlapped states. The features are extracted using the color and shape descriptors, and are classified into TB bacilli and non TB bacilli. After classification, the TB objects are separated from non TB objects and are counted to find the level of tuberculosis infections as shown in Figure \ref{fig1}.

\subsection{Image acquisition}
\begin{figure}[h!]
\centering
\includegraphics[width=10cm, height=7cm]{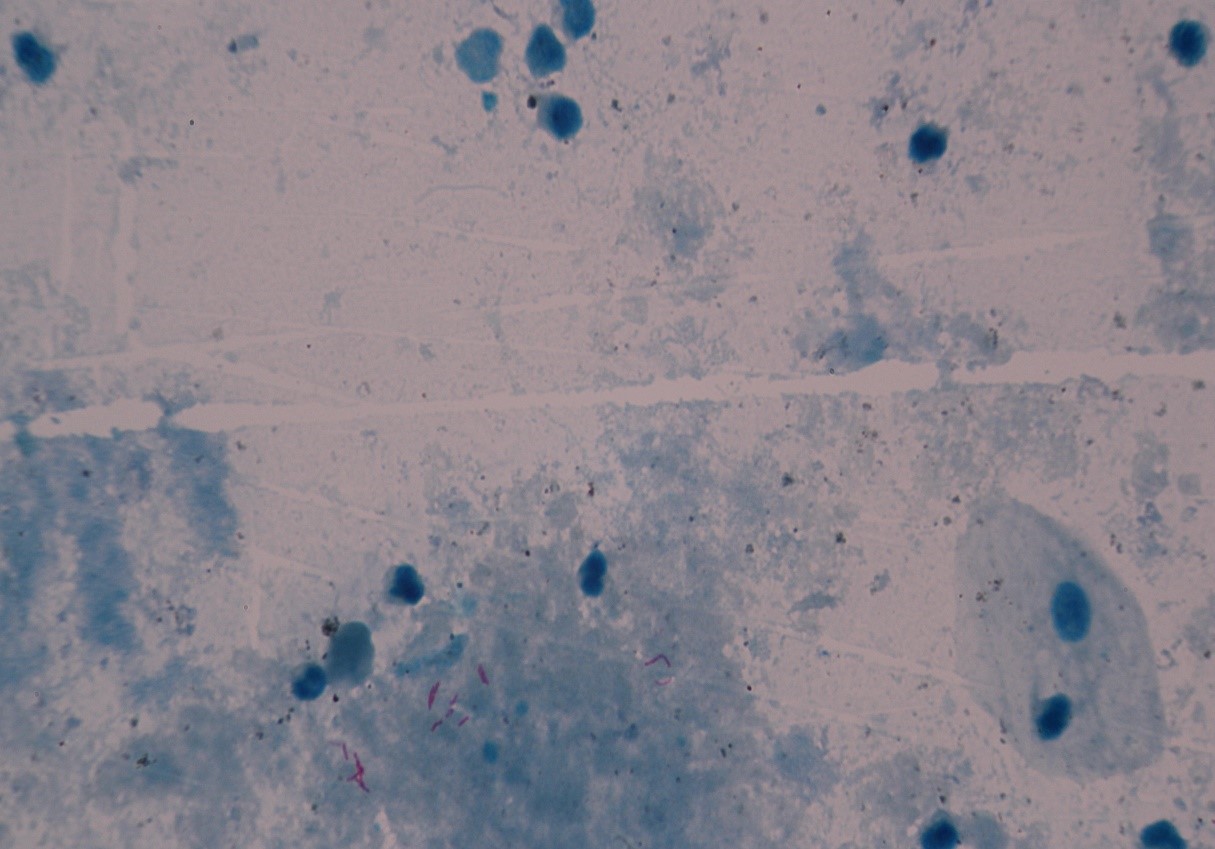}
\centering
\caption{Sample conventional microscopic image captured at 100x magnification containing TB bacilli in purple color after Ziehl-Neelsen staining}
\label{fig2}
\end{figure}

Olympus CX-21i binocular microscope is used for the examination of smear. The ZN stained sputum smears are placed on the microscopic stage for AFB examination. A high definition canon 1200D DSLR camera is attached to the microscope for digitizing the image. Sputum smear slides are then focused for AFB under 100x magnification. On focusing, the Field of views are captured using a camera attached to the ocular lens in a microscope. The captured 24 bit image is of size $5184 \times 2912$ resolution with 72dpi and a focal length of 50mm as shown in Figure \ref{fig2}. These digitized images are then processed for TB bacilli identification.

\subsubsection{Image enhancement}

The digitized microscopic image is then pre-processed for TB bacilli detection. The acquired image is a 24 bit RGB color image with high resolution. The RGB image has the red, green and blue color component for each pixel. A pixel color is determined by the combining intensities of red, green and blue components. For enhancing the image quality, RGB planes are separated and contrast stretching is done for individual RGB planes \cite{inbook}. The brightness and the contrast of the image is enhanced by contrast stretching. The pixel value is quantized to a new pixel value by a pre-specified function. Contrast stretching is a linear mapping function which applies auto-scaling. The mapping function for contrast stretching is illustrated in Equation 1.
\begin{equation}
	p_k=\frac{(max-\ min)}{f_{max}-f_{min}}(q_k-f_{\min)}+min 
\end{equation}
Referring to the above Equation 1, $f_{max}$, $f_{min}$ are the maximum and minimum pixel color level in an input image. Input color level is denoted by $q_k$ and output pixel color level is denoted by $p_k$. After contrast stretching, the planes are concatenated to obtain a quality image.

\subsection{Segmentation process of TB bacilli}

\begin{figure}[h!]
\centering
\includegraphics[width=10cm, height=7cm]{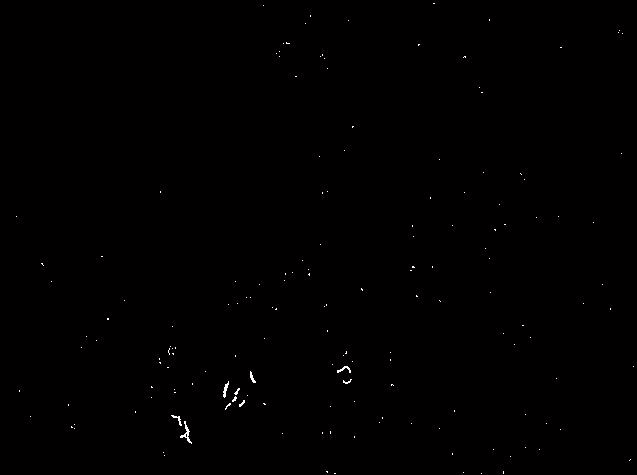}
\centering
\caption{Coarse level segmented image with TB bacilli and debris}
\label{fig3}
\end{figure}

The enhanced image is then segmented into foreground and background using image thresholding. Here, the TB bacilli object, which appears in red color were separated from the bluish background. To segment the TB bacilli, coarse and fine level segmentation are done to separate the bacilli objects from background. In coarse level segmentation, TB bacilli object pixel is marked as foreground and are compared to a threshold value $(T)$ to separate it from the background as in Figure \ref{fig3}. The process is repeated till the average of each set of pixels is computed (i.e. foreground and background pixels).

In fine level segmentation, adaptive local thresholding techniques based on the local values of pixels and their neighbors were computed. The coarse segmented image is taken as input and local threshold of each pixel intensity is computed with respect to the neighborhood pixels in fine segmentation. 
Here, Sauvola’s technique is used for computing the adaptive local thresholding \cite{10.1007/s10032-013-0209-0}. Let $g(x,y)$ be the pixel intensity at a location $(x,y)$ as described in Equation 2.

\begin{equation}
  O \left(x,y\right)=\begin{cases}
    0, & \text{if g(x,y)$\leq$ t(x,y)}.\\
    255, & \text{otherwise}.
  \end{cases}
\end{equation}

A local threshold value $t(x,y)$ is computed for each pixel using the mean $m(x,y)$ and standard deviation $s(x,y)$ of all pixel intensities around the pixel in a $W \times W$ window. 
\begin{equation}
	t\left(x,y\right)=m(x,y)\left[1+k\left(\frac{s(x,y)}{R}-1\right)\right]                                                   
\end{equation}

Where $R$ in Equation 3, denotes the maximum standard deviation value of $s(x,y)$ and parameter $k$ is assigned a positive value in range $[0.2,0.5]$. The local values to the mean $m(x,y)$ and standard deviation $s(x,y)$ comply threshold from the contrast value of the neighborhood pixels. The high contrast regions produce $s(x,y)\approx R$ which implies $t(x,y)\approx m(x,y)$. When the difference between the local neighbourhood pixels is low, then $s(x,y)\neq R$.The value of threshold in the local window $W \times W$ is controlled by $k$ parameter. If the threshold is lower from the local mean $m(x,y)$, then the value of $k$ is higher. For a $N \times N$ size image, the computational complexity to compute the mean $m(x,y)$ and standard deviation $s(x,y)$ is of $O(W^2N^2)$. Sauvola et al. proposed an efficient way to compute the threshold for each $N^{th}$ pixel and then to use the interpolation for rest of the pixels \cite{10.1007/s10032-013-0209-0}. Thus the computation speeds up for determining the threshold.

\begin{figure}[h!]
\centering
\includegraphics[width=10cm, height=7cm]{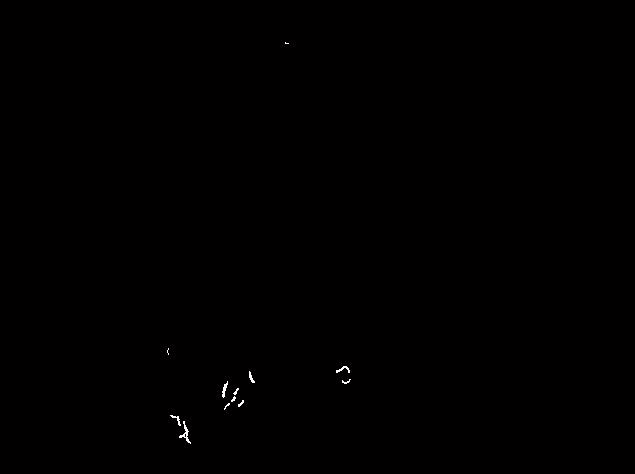}
\centering
\caption{Extracted TB bacilli features after elimination of non-TB bacilli debris}
\label{fig4}
\end{figure}

After the fine and coarse level segmentation, both the segmentation results are merged to obtain a foreground with TB bacilli objects and some debris. To remove the debris from the segmented image, 8 pixel connected component analysis is used.  On detecting the objects, the set of connected pixels were analyzed. These pixels do not define a finite boundary and constitutes an object, which can be a TB bacilli or non TB bacilli.

\subsection{Shape descriptors for TB bacilli detection}
The morphological operations are done on the segmented image with some shape descriptors to extract the feature as shown in Figure \ref{fig4}. The tuberculosis bacillus is of rod shaped, which has a defined shape and size parameters for identification. To identify TB bacilli and non TB bacilli, geometrical and shape features are required. These geometrical features are extracted using the following shape descriptors \cite{article15}.

\begin{itemize}
    \item Area (A) and Perimeter (p): The area estimates the total number of pixels corresponds roughly to a particular object. Perimeter refers to the continuous occurrence of pixels along, boundary of the object. 
    \item Circularity or Roughness: The circularity refers to the roundness of object, which ensures the object not being too long or square. Circularity uses features such as cylinders, spheres and cones.
    \item Major axis length: The major axis length is the longest diameter, running through the center and it touches the extensive points in the perimeter.
    \item Convex hull perimeter: The convex hull have set of X points, which bounds the subset of the plane with the X subsets. The convex hull may contain points that are not in the subset. Hence, remove the point from the subset that is not a hull and replace it with a point to the interior of hull to find the object shape.
    \item Eccentricity: The eccentricity is referred to the longest contour of a bacilli shape to the longest perpendicular contour.
\end{itemize}

\section{Discussion}
\begin{figure}[h!]
\centering
\includegraphics[width=10cm, height=7cm]{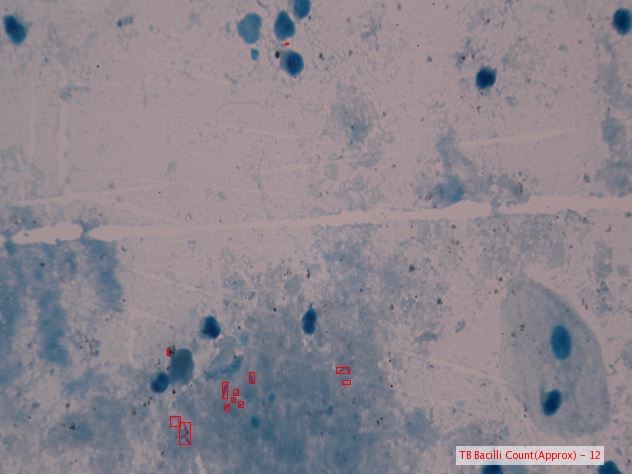}
\centering
\caption{Detection and counting of TB bacilli from the infected field of view}
\label{fig5}
\end{figure}

The proposed methods and techniques were applied to the Ziehl-Neelsen stained microscopic images. Few samples of images were taken to find the robustness of the methods. The foreground, which has both the TB bacilli and non TB bacilli is segmented from the image and feature descriptors are used to eliminate the non TB bacilli. In case of clustered or overlapped bacilli in the image, proposed method counts the clustered set of bacilli as one, which decreases the sensitivity of diagnosis. However, the proposed technique produces a good result for images with single TB bacilli and counts the total number of bacilli per field of view in an effective way 16 as shown in Figure \ref{fig5}. The overall sensitivity and specificity of detecting non-overlapped TB bacilli is 98.7\% and 94.3\%.

\section*{Conclusion}
An automated TB bacilli identification technique is proposed to find the severity of infections. The study is conducted on segmentation techniques, color and shape descriptors to separate the TB bacilli and non TB bacilli. The RGB pixel characteristics, adaptive thresholding, feature extractions are done to identify the TB bacilli. Satisfactorily, completing the previous processes, the identified TB bacilli are counted to report the severity. The severity is classified into four classes, TB negative, scanty, 1+ and 2+. Subsequently, after identification and counting, the status of the patients with their severity was updated in the transaction database for clinical records.

\section*{Acknowledgement}
This research was supported by Pondicherry Institute of Medical Sciences (PIMS), Pondicherry, India. The authors also wished to show their gratitude to Dr. Anil Jacob Purty, Registrar, PIMS for sharing the ZN stained sputum smear specimen during the course of research. 

\bibliographystyle{unsrt}  
\bibliography{references}  



\end{document}